\documentclass{IEEEcsmagModified}

\usepackage[colorlinks,urlcolor=blue,linkcolor=blue,citecolor=blue]{hyperref}

\usepackage{amsmath}
\usepackage{upmath}
\usepackage{comment}
\usepackage{tabularx}
\usepackage{caption}
\usepackage{subcaption}
\usepackage{booktabs}
\usepackage{float}
\usepackage{breakurl}

\jvol{XX}
\jnum{XX}
\paper{XX}
\jmonth{}
\jname{}
\pubyear{2024}

\begin{document}

\title{LLMs for Cyber Security: New Opportunities}

\author{Dinil Mon Divakaran}
\affil{A*STAR Institute for Infocomm Research}
\author{Sai Teja Peddinti}
\affil{Google}

\begin{abstract}
Large language models (LLMs) are a class of powerful and versatile models that are beneficial to many industries. With the emergence of LLMs, we take a fresh look at cyber security, specifically exploring and summarizing the potential of LLMs in addressing challenging problems in the security and safety domains.

\end{abstract}
\maketitle


\section{Introduction}

Large Language Models (LLMs) are creating a transformational impact in the space of science and technology, giving rise to a wide variety of new applications for various services across diverse industry verticals. Their capability to comprehend and, in particular, to generate contents, represents a paradigm shift that is reshaping the way we interact with computers, leading to the development of numerous innovative applications. Today, LLMs are able to generate text, images, and videos; there are LLM applications that hold conversations with humans, translate between languages, explain and write code, resolve programming bugs, and so forth. 

LLMs generally are based on a transformer architecture that uses self-attention mechanism to efficiently learn long-range dependencies of tokens (words or sub-words) in a sequence of data (e.g., a sentence). This has allowed transformer models to not only improve upon previous sequence models such as RNNs (Recurrent Neural Networks), but also to train large models of billions and even trillions of parameters on datasets of massive sizes. Importantly, the {\em pretraining} of an LLM is unsupervised, removing the burden of labeling large datasets. Like other generative models, LLMs fundamentally aim to recreate data they are trained on.  Using these properties, pretrained LLMs have been used to generalize across many tasks, often by fine-tuning on small amounts of labeled data. GPT-4, Gemini, Llama~2, Mistral, Falcon, OLMo (Open Language Model), etc., are some of the well-known LLMs today, while new ones are being built at a rapid pace. Examples of downstream tasks include language translation, sentiment analysis, domain-specific chatbot conversation, text based image/video generation, assistive medical diagnosis, etc.

Unsurprisingly though, such a compelling technology can be put to dual use. An LLM is fundamentally a probabilistic model, which learns to make predictions based on the massive datasets that it has been trained on; and thus, it is only reasonable that the model may not consistently generate factually accurate, benign, or positive outputs, even if trained to do so. This inherent characteristic can be exploited, e.g., via prompt injection attack (discussed later), by malicious actors for various purposes. We refer the reader to the `NIST Trustworthy and Responsible AI report (2023)'~\cite{nist2023report}, 
for a detailed taxonomy of adversarial machine learning (ML) in the context of both conventional ML as well as LLMs.

There are ongoing efforts to mitigate the risks due to LLMs. 
Companies such as OpenAI (\url{https://openai.com/safety}), Google (\url{https://safety.google/cybersecurity-advancements/saif/}), 
Meta (\url{https://ai.meta.com/responsible-ai/}), Microsoft (\url{https://www.microsoft.com/en-us/ai/responsible-ai}), etc. have frameworks for developing safe and responsible AI systems. In fact, many of the firms also focus on {\em red teaming} LLMs, to proactively investigate and identify vulnerabilities of LLMs, e.g., to detect adversarial prompts that can generate harmful or malicious responses. 
In 2023, Microsoft, Anthropic, Google, and OpenAI launched the {\em Frontier Model Forum}~\cite{frontier-model-forum-blog}
to support best practices to mitigate risks, advance research on AI safety and security, as well as facilitate information sharing among companies and governments. Similarly, companies formed a {\em C2PA coalition}~\cite{c2pa}
to create an open technical standard that will aid in the ability to trace the origin of different types of generated media.  Lastly, governments across the world are also working on regulatory frameworks for AI, to protect AI users and user privacy (among others). It is worth noting that, governments are encouraging global collaborative efforts to tackle AI vulnerabilities and security risks (e.g., refer the U.S Executive Order on AI~\cite{US-exec-AI-ACT-2023},
and the European Union's AI Act~\cite{EU-AI-Act-2023}).

\subsection{New opportunity to address cyber security problems}
\begin{figure*}
\centerline{{\includegraphics[scale=0.45]{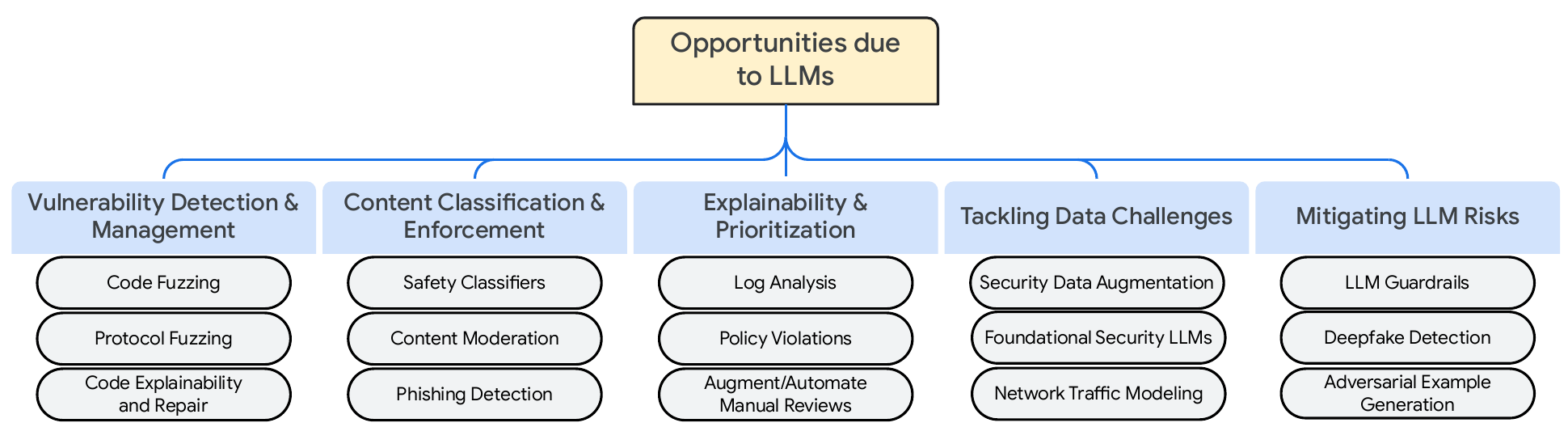}}}
\caption{LLMs offer versatile solutions to address a wide range of cyber security challenges.
\label{fig:LLM-opportunities}}

\end{figure*}

We now turn to the main focus of this article and discuss the new opportunities LLMs present in addressing security and safety challenges that users today face in the digital world.
The cyber security domain has already started to see the benefits of utilizing LLMs for addressing some of the important problems in the domain, and we summarize some of these recent advancements. These efforts can be broadly categorized into five themes described below. Refer to Figure~\ref{fig:LLM-opportunities} for an overview.

\section{LLMs for Vulnerability Detection and Management}

Today, there are multiple LLM-based tools that are being built to help with code development. Generating code based on natural language description has the promise to transform the software development domain. Devin AI, GitHub Copilot, IBM's watsonx, Amazon CodeWhisperer and Codeium are some of the emerging AI code assistants. They perform advanced tasks such as code generation and completion, code repair, code refactoring, and code explanation. Besides lowering the entry barrier for software development, these code assistants help in reducing bugs in software development process. For instance, propagating changes in variable type {\em automatically}, although appears simple, is a particularly useful feature that helps developers.

The number of CVEs published has been increasing over the years and approached close to $29,000$ in 2023~\cite{cve-metrics}.
A 2024 report from Synopsys states that the proportion of codebases that have high-risk vulnerabilities---including exploited vulnerabilities---increased from 48\% in 2022 to 74\% in 2023~\cite{synopsys-2024}. 
% https://investor.synopsys.com/news/news-details/2024/New-Synopsys-Report-Finds-74-of-Codebases-Contained-High-Risk-Open-Source-Vulnerabilities-Surging-54-Since-Last-Year/default.aspx
Software vulnerabilities lead to system failures, and malicious actors target the vulnerabilities to launch cyber attacks. 
While AI-generated programs are not perfect and could also be vulnerable, they hold promise in comparison to human developers---an empirical study by Asare {\em et al.} demonstrates less vulnerabilities introduced by AI code assistants than humans~\cite{Copilot-study-2023}. 
Another user study assessing LLM-assisted coding of 58 students also indicates low security risk due to LLMs~\cite{sandoval2023lost}.
Besides, researchers are studying how LLMs could be utilized to not only detect vulnerabilities~\cite{LLM4vuldetection-results-2024}, but also to {\em automatically} repair code vulnerabilities~\cite{zero-shot-vul-repair-LLM-S&P-2023,zhang2024autocoderover}. Indeed, the results from~\cite{zhang2024autocoderover} are promising: the proposed solution AutoCodeRover resolved 67 GitHub issues, each taking less than 12 minutes; this is much faster than the time taken by human developers (more than two days on average). 
Google shared that its Gemini model helped successfully fix 15\% of bugs discovered by their sanitizer tools, resulting in hundreds of bugs patched~\cite[Section~5]{googledefendersdilemma}. 

Furthermore, given that LLMs are pretrained on vast amounts of online data including source code and RFCs of protocols, new research illustrates the potential of LLMs in guiding protocol and code fuzzing for vulnerability discovery. The protocol fuzzer \textit{ChatAFL}~\cite{chatafl-2024}
capitalizes on the understanding of the RFCs the LLMs have. 
\textit{Fuzz4All}~\cite{fuzz4all} is a universal code fuzzer that can target many different input languages and many different features of these languages, and it has been shown to discover bugs and vulnerabilities in software systems. 
Also, competitions such as the \textit{AI Cyber Challenge}\footnote{https://aicyberchallenge.com/},
a two-year competition announced in late 2023, organized by DARPA in collaboration with others to design and develop AI-based solutions to secure code, have given momentum to this line of research.  

\section{LLMs for Content Classification and Enforcement}
LLMs are being leveraged to augment or automate several general purpose security/safety classifiers, some of which are described below. 

\paragraph{Safety Classifiers for Policy Enforcement:}
Toxic contents are on the rise on online platforms. Hate speech, harassment, cyber-bullying, etc. adversely affect users of all communities, and in particular underrepresented groups. 
The complexity of this socio-technological problem is amplified by the multilingual nature of communications, the use of evolving lingo, emojis, styles, and so forth.
One of the well-known classifiers for toxic content detection that is used by developers and publishers is Google Jigsaw's Perspective API\footnote{https://perspectiveapi.com/}. 
The collaborative team has been publishing tools  and data, besides improving the model capabilities. There are also a number of ML models proposed in the literature to address this issue. 

Despite the active research in toxic content detection, the scarcity of large-scale, high-quality data impedes research. However, LLMs pretrained on massive data offer a promising direction. As noted previously, LLMs have the capability to solve downstream tasks with a small number of labeled samples, or even without fine-tuning. Indeed, He {\em et al.} show that, with prompt learning--giving a few examples at an LLM's prompt, pretrained LLMs are able to achieve  better performance than models trained specifically for toxic content detection~\cite{LLM-prompt-toxic-S&P-2024}. 
That said, the problem is far from being solved. We have to develop solutions that extend beyond text analysis to detect toxicity in various media formats, including images, audios, videos, and obfuscated messages. Continued research in the field of LLMs, aimed at enhancing their capability to perform on tasks across diverse content formats, holds the potential to offer new solutions for combating toxic content in online platforms. 

Another area where LLMs are useful is content moderation. Content safety policies often evolve too frequently to catch-up with the different types of threats emerging online. LLM's \textit{zero-shot} capabilities are immensely valuable in quick enforcement of these evolving policies, or for reducing labeling costs when creating annotated datasets for training down-stream ML models. Kumar {\em et al.}~\cite{kumar2024watch} show that LLMs (such as GPT-3.5) are effective at rule-based moderation for many Reddit communities, achieving performance close to human moderators for some communities. This early result motivates exploring LLM use for content moderation in other settings.

\paragraph{Phishing Detection:}
Phishing is one of the most common cyber attacks in recent times. Attackers craft and send phishing emails to victims, often including text, image (e.g., brand logo) and a URL to a phishing website. Phishing emails can be targeted to specific individuals (say, a person in the Finance department of a company), and links to phishing websites are also distributed via social media, chats, SMSes, etc. This also presents multiple options for solution development. For example, specific phishing detection solutions are integrated with email and SMS gateways. Also, threat intelligence services get URLs from various sources and analyze them using standalone services. A popular service is VirusTotal, which utilizes more than 70 URL-analyzing engines from cyber security vendors and provides aggregate results to users. Despite these protections in place, many (carefully crafted) phishing emails are evading these scanners and reaching users' mail boxes. 

{\bf Phishing emails.} Over the years, phishing email solutions have evolved from relying solely on rules and signatures to the use of ML models to automatically learn patterns of phishing emails. Recently, we also see the use of LLMs for addressing this threat. 
The phishing detection system D-Fence~\cite{d-fence-2021} uses the LLM BERT to generate the embeddings of texts in emails, and subsequently uses the embeddings along with other features to train a  model for classifying emails as either phishing or benign. 
Koide {\em et al.}~\cite{koide2024chatspamdetector} created \textit{ChatSpamDetector}, that utilizes LLMs to detect phishing emails and  obtain detailed reasoning for the phishing determination. This system is shown to outperform existing baseline detection systems, does not require continuous updates to the detection models and block lists like in existing spam filters, and the generated rationales assist users in making informed decisions when handling suspicious emails.

{\bf Phishing webpages.} A well-known approach to detecting phishing webpages, called {\em reference-based approach}, is to compare the logos on a given webpage to a known {\em reference} set of logos of 
popular brands (e.g., Paypal, Amazon, etc.)~\cite{phishpedia-usenix-sec-2021, phishintention-usenix-sec-2022}. The basic idea in reference-based approach is that, if a webpage contains a well-known brand's logo (e.g., Paypal's) but has a different domain name, then it is a phishing page. The state-of-the-art solution, Phishpedia~\cite{phishpedia-usenix-sec-2021},
trains an object-detection model to detect the logos on screenshots of webpages and a Siamese model to identify the brand of a detected logo. 
Making advancement in this direction, in~\cite{li2024knowphish}, the authors use an LLM to extract brand information from the text present in the HTML pages as well as  to detect whether the webpage solicits user credentials (login/password). This approach enables the detection of phishing pages with or without the presence of logos, expanding the capabilities of existing reference-based detection methods (such as Phishpedia).

Another potential research direction for phishing webpage detection is, training or fine-tuning an LLM pretrained on large-scale website dataset for HTML understanding and semantic information extraction~\cite{HTML-understanding-2023}. 
Similar models, and even multi-modal LLMs that can take texts and images as input, could be utilized for building classifiers that detect phishing webpages. 
As a concrete example, a model pretrained on large-scale webpage dataset can be fine-tuned using benign and phishing pages to develop a phishing webpage classifier. To reduce the high maintenance costs of such LLM-classifiers,  \textit{distillation} techniques can be used to transfer learnings to smaller sized models for wide-scale deployment. 

The use of Large Language Models (LLMs) in combating phishing attacks is gaining research traction. However, significant challenges remain. Firstly, attackers have the ability to perturb logos of brands they use in their phishing attacks, and thereby evade logo-based or reference-based phishing detectors~\cite{phishing-ESORICS-2023}. Secondly, legitimate logos from popular companies used in single sign-on (SSO) or advertisements can trigger false positives~\cite{phishpedia-usenix-sec-2021}. Finally, not all phishing attacks rely on logos~\cite{li2024knowphish}.

\section{LLMs for Explainability and Prioritization}

LLMs, with their natural language interface and the ability to work with data in multiple modalities (text, images, videos, code, etc.), can help with understanding diverse data. Newer LLMs, such as Google's Gemini Pro 1.5 and Anthropic's Claude 3 Haiku, boast extremely large context windows of more than 100,000 tokens, enabling them to digest and summarize large amounts of data. These capabilities have opened up new avenues of utilizing LLMs for data explainability, summarization, and for automating or augmenting human reviews.

\paragraph{Explainability:}

Enterprises deploy security solutions from one or more vendors to protect their endpoints. To gain high visibility, modern security solution providers gather detailed data from processes, network connections, applications, file/registry accesses, etc., thus resulting in humongous logs. SentinelOne Singularity, CrowdStrike Falcon and Trend Micro Apex are examples of commercial EDR (endpoint detection and response) solutions. Besides the logging capability, EDR solutions also come with a set of rules to detect malicious patterns of known malware. 
Similar problem also exists in the cloud and distributed computing systems. For example, the promising microservice architecture that helps to scale up resources as required for an application, also comes with threats due to insecure packages, misconfigured authentications, etc. The large attack surface exposed due to the distributed nature of the architecture makes it all the more relevant to log information and analyze them in real-time for timely detection of anomalies and attacks. 

As traditional approach of writing rules to match malicious patterns neither scales nor achieves high detection accuracy, security researchers are developing ML models that train on huge amounts of process/audit logs to detect suspicious behaviors. However, this creates another challenge---the detected patterns from the endpoints need to be investigated by security analysts to take the appropriate mitigation steps. Besides, ML models also raise false positives; and a high number of patterns that need to be investigated leads to {\em alert fatigue}, which in turn results in missing out high-risks threats and attacks. Cyber defenders' burn-out is a known chronic problem~\cite[Section~3]{googledefendersdilemma}. LLMs are currently being used to explain the detected patterns, to make it easier for an analyst to decide quickly. For example, {\em HuntGPT}~\cite{ali2023huntgpt} is a specialized intrusion detection dashboard that uses LLMs to discern patterns in network traffic and deliver detected threats in an understandable format. Powered by GPT-3.5-turbo, the system achieved more than 80\% success rate at the CISM (Certified Information
Security Manager) Practice exams, showing promise in guiding security decisions.  
Other examples from a recently published Google report~\cite{googledefendersdilemma} include the following. i)~The Google Detection \& Response teams have leveraged Gemini LLM for natural language querying and automatic summarization of alerts data, and have seen a 51\% time savings and higher quality results in incident analysis. ii)~Google Cloud's SecLM, a security-specific LLM, facilitates analysts to conversationally search and interact with security events, provides explanations for complex attack graphs, and even recommends mitigations. Similarly, VirusTotal Code Insight explains what a potentially malicious Powershell code is doing~\cite{VT-insights-2023},
and solutions such as CrowdStrike's Charlotte AI, Google Cloud's DuetAI, and
Microsoft's Security Copilot
also aim to empower security analysts in their threat hunting process. Such assistive solutions can help even non-expert security analysts to detect, investigate, and respond to cyber threats with confidence. 

Performing content moderation across online platforms has very similar challenges, where human reviewers have to investigate a multitude of (ML or user) flagged posts for policy violations. Kumar {\em et~al.}~\cite{kumar2024watch} show that the reasoning capabilities of LLMs are immensely useful in providing explanations and in identifying the specific rules being violated by the policy violating posts, making LLMs a valuable aid for humans performing content moderation.

\paragraph{Prioritization:}
LLMs are also very useful in automating or augmenting manual reviews, and help reduce a reviewer's  fatigue when sifting through detected security incidents or flagged online content. They help evaluate the veracity of identified incidents or policy violations, automating decisions in clear cases and triaging/escalating high risk, complex, or borderlines cases to help focus engineering/expert resources efficiently. For instance, Qiao {\em et~al.}~\cite{Qiao_2024} employed LLMs to scale up content moderation in Google Ads. They were able to reduce the number of manual reviews by more than 3 orders of magnitude while achieving a 2x recall compared to a baseline non-LLM implementation. 

Automated decision making of LLMs also helps reduce exposure of human reviewers to harmful content, thereby enhancing their mental well-being. Puentes {\em et~al.}~\cite{puentes2023guarding} propose a Large Language Model (LLM) that analyzes and classifies the information received in reports on sextortion, sexting, grooming, and sexual cyberbullying. Their system even efficiently forwards the
reports to competent authorities, and reduces the exposure of analysts  to harmful contents.

Despite LLM's strengths in content summarization, explanability, and automation, they are known to be prone to hallucinations---where they generate responses that are factually incorrect, nonsensical, or disconnected (from inputs). Research focusing on `grounding' the LLMs to the provided data can alleviate these concerns.

\section{LLMs for Tackling Data Challenges}

Building highly accurate ML models for security and safety use cases requires large labeled datasets. In the domain on cyber security, there are two challenges in obtaining quality datasets for training models. 

\begin{itemize}
    \item Labeling Cost: As in many domains, labeling is a costly task requiring human effort. To develop ML models for solving security problems (such as detection of network attacks,
    malware detection via static and dynamic analyses, 
    etc.) requires large labeled datasets. While the research community publishes data once in a while, they are limited in size, may contain artifacts (e.g., malicious datasets for network attacks and endpoint logs for malware analysis are often generated via emulation in a controlled environment), or may be obsolete. 

    \item Data Privacy and Retention: Another challenge in obtaining real-world dataset is the risk of  leaking sensitive or confidential information.
    Consider email data (required for phishing detection), social media data (required for content moderation), network traffic, etc., where there is risk of privacy leak. Even though privacy-preserving transformation of data can be performed before making the dataset available for research, the risk of leak is so high that, such real-world datasets are only available to researchers of the corresponding firm who `owns' the data -- which also affects the reproducibility of the research works in this domain.
    On the other hand, to provide privacy guarantees, companies often employ retention timelines when storing user data, that indicate how long the data can be stored and used. 
    Often these retention requirements also get applied to the manually annotated training data, when it is derived from user data. 
    For instance, consider the case of a toxicity model trained on social media data. Based on the policy that {\it a user's data would be deleted from the social media website within a week after they delete their account}, the toxicity model would start forgetting patterns seen across deleted users' data. For model performance benchmarking over time and to avoid forgetting patterns observed in old data, permanent access to annotated training data is necessary. 
\end{itemize}

Given the above challenges, LLMs are being explored for data augmentation needs. Data augmentation techniques help with diversifying training examples without the need for additional data collection or labeling. For instance, Lee {\em et~al.}~\cite{lee2024llm2llm} have proposed \textit{LLM2LLM}, an iterative data augmentation strategy to enhance a small-seed dataset, and have demonstrated that this reduces dependence on labor-intensive data curation while simultaneously achieving improvements over regular fine-tuning in low-data regime tasks. Others are leveraging LLMs for augmenting training datasets in new languages (to enhance cross-lingual performance of base models~\cite{whitehouse2023llm}), or are exploring synthetic data generation approaches for completely skipping training data annotation~\cite{mengICML23}. To avoid any privacy leaks, LLMs are also being fine-tuned on sensitive datasets in a `differentially private' way~\cite{shi2022just}. 

\paragraph{Traffic Modeling for Network Security:}

However, data augmentation alone isn't sufficient. Consider network traffic analysis for detection of various threats, anomalies and attacks. Years of research works have led to the development of numerous statistical, ML and data mining algorithms for network security tasks, such as detection of bots, C\&C channels used for communication between attacker and compromised hosts, low-rate DDoS attacks, password-spraying attempts, generic anomalies, etc.~\cite{GEE-2019, detecting-bots-GAN-S&P2020, low-quality-nw-NDSS-2024}. Each of these tasks require large amounts of labeled data with minimum noise for training and evaluating ML models; but large-scale, high-quality data is not available (e.g., see~\cite{low-quality-nw-NDSS-2024}). For example, to train a model for detecting bot traffic to an e-commerce website, the dataset has to have hundreds of thousands of labeled network requests that are made by both bots and legitimate users~\cite{detecting-bots-GAN-S&P2020}. Yet, it is arguable whether such a dataset helps in building models that can generalize well, given data can come from different operating systems, browsers, locations, etc. 
Therefore, to generalize, and even to sustain model by retraining, such an e-commerce entity would have to label network traffic regularly. 

The advancements in LLM development present a new opportunity to train domain-specific foundational models in an unsupervised way. In the case of network traffic analysis for detecting threats and attacks, a network-specific foundational LLM that learns network `conversations' (e.g., requests and responses) can be trained using openly available real-world network traffic datasets. CAIDA and MAWI, for example, regularly publish network traces for research purposes; while being massive in size, they are mostly unlabeled. But these unlabeled datasets can be utilized for training a network LLM. Such a foundational LLM can then be fine-tuned for multiple downstream tasks, such as botnet detection. 
Although fine-tuning is a supervised approach, it typically requires only small amounts of labeled data, thereby decreasing labeling costs significantly. 
The network research community is witnessing active discussions in this direction, of training an LLM that learns network communication language (see ACM HotNets 2022~\cite{hotnets-2022} 
and 2023~\cite{hotnets-2023} proceedings).

\section{LLMs for Mitigating LLM Risks}

With their generative capability, LLMs have lowered the entry barrier for cyber criminals. Phishing emails, tailored to specific roles or individuals, can be generated using LLM applications such as ChatGPT~\cite{exploiting-LLM-dual-use-ICMLworkshop-2023}. 
Researchers at CyberArk outlined how to generate polymorphic malware; the malware runs with ChatGPT API generating new payloads and malicious modules as and when required to evade detection~\cite{cyberark-polymorphic}. 
Security researchers have already discovered generative AI tools in the dark web marketplaces that help attackers with their cyber criminal activities; examples include FraudGPT~\cite{fraudgpt} and WormGPT.
And attackers are exploiting the capability of LLMs to generate highly realistic and convincing images, videos, and audio to create {\em Deepfakes}~\cite{deepfakes-survey-2021}. 
Deepfakes are already being used for unethical and malicious purposes such as spreading misinformation, generating fake news, and defaming individuals. 
Microsoft lists a number of threat actors that have adopted generative AI tools to launch recent attacks~\cite{MS-intel-AI-actors-2024}.

While the above attacks are not novel {\em per se}, their proliferation is enabled by LLMs, specifically due to a new attack vector of LLMs, namely {\em prompt injection}. 
In this attack, an attacker exploits the ability to query LLM models through well-defined APIs and interfaces to either {\em extract} sensitive information (such as application product keys), 
or enable scope for other threats such as remote code injection. For instance, prompt injection attacks enabled researchers from Juniper Networks to trick ChatGPT to generate malware code~\cite{juniper-mal-code}, 
and the security vendor Bitdefender to solve a CAPTCHA~\cite{bitdefender-Bing}. 
The attack surface increases when an LLM is extended with data sources to provide more up-to-date information via retrieval augmented generation (RAG), thereby blurring the line between instruction and data~\cite{indirect-prompt-AISec-2023}.
An example is of an attacker sending an email with malicious instructions that are automatically fed to an LLM application meant for detecting spam or phishing emails, but then inadvertently follows the attacker's instructions.
Prompt injection attack is recognized as the top LLM related attack by OWASP~\cite{owasp-top10-LLM};
and they are of particular concern when new applications interface with an LLM for automated responses~\cite{LLM-app-attack-2023}. 

To negate the above mentioned LLM risks and vulnerabilities, there is also research studying and deploying a multitude of security risk mitigation strategies, including defining and applying strict policies for moderating the input and filtering the output.
One approach is to have safeguard checks and controls, also termed as {\em guardrails}, in place. 
For example, safety filters in text-to-image models, such as DALL-E 2 and Midjourney,  prevent generating not-safe-for-work (NSFW) content. {\em Llama Guard}~\cite{inan2023llama} from Meta is an LLM trained to classify an LLM prompt or a response as safe. 
In a recent work~\cite{kumar2023certifying}, researchers have shown that fine-tuning a pretrained DistilBERT model on labeled safe and harmful prompts is more effective in detecting harmful prompts than safety filters of Llama-2, due to fine-tuning on the specific task. 
There are also independently developed guardrail solutions focusing on a specific data type and task, such as unsafe image detectors (see, e.g.,~\cite{LAION-AI}), and LLM firewalls that can block malicious prompts and sensitive/harmful outputs~\cite{LLM-firewall}. 
Developing these guardrails is a challenging and ongoing effort, as they have to catch up to different models, applications and evolving policies. 
Besides, recent research shows that bypassing LLM defenses is possible today through prompt injection attacks even when the LLMs are {\em safety-aligned}~\cite{ba2023surrogateprompt, jailbreak-NDSS-2024}.

Countering the challenge of exploiting AI-generated contents (deepfakes) for fraudulent purposes is an active area of research within the AI domain, and one of the interesting research directions is to add watermarks
to contents generated by LLMs. A noteworthy contribution in this area comes from Kirchenbauer {\it et al.}~\cite{kirchenbauer2023watermark}. Their work proposes a watermarking framework that i)~generates a watermarked text without requiring the LLM to be retrained, ii)~enables subsequent identification of watermarked text with negligible false positives. 

ML-based defense solutions are susceptible to evasion attacks. 
For example, attackers could generate {\em perturbed} logos of reputed brands in their phishing campaigns to persuade human users into divulging their credentials. The perturbations are such that they are imperceptible to the human eye while being effective in evading phishing defense solutions based on logo identification (using ML models)~\cite{phishing-ESORICS-2023}. 
A well-studied approach to counter such evasions is \textit{adversarial training}, where training with adversarial examples (e.g., perturbed logos) can enhance the robustness of defense models against evasion attacks (see~\cite[Section~7]{phishing-ESORICS-2023}). 
With their generative capabilities, LLMs are being leveraged to automate the generation of adversarial examples with little human effort~\cite{guo2024autoda, struppek2024exploring}. These adversarial examples can then be incorporated into adversarial training to build robust models to defend against threats and attacks. 

\section{Key takeaways}

There is an inherent asymmetry between the attackers and defenders in the cyberspace, popularly referred to as ``Defender's Dilemma'', which states that it is sufficient for an attacker to succeed once but a defender must be successful in protecting at all times~\cite{googledefendersdilemma}. Machine Learning and Artificial Intelligence (AI), and specifically Large Language Models, have the potential to tilt the scales of cyberspace to give the defenders an advantage over the attackers. The emergence of LLMs presents an opportunity to reimagine how we approach and solve cyber security challenges, enabling the development of innovative solutions by leveraging the capabilities of these powerful models. There are early works indicating that LLMs are helpful in this regard -- in defending against software vulnerabilities, phishing attacks, network threats, moderating toxic content on social networks, etc.
A recent MIT study has shown that inexperienced workers stand to gain the most from generative AI solutions, such as LLMs, while skilled workers gain incremental benefits~\cite[Section~5]{googledefendersdilemma}. In other words, generative AI solutions are democratizing security expertise for everyone and are being termed as the ``great equalizer''. Organizations without much security expertise are leveraging AI assistive solutions for improving their security postures. Similarly, experiments are being carried out to evaluate the effectiveness of LLMs in succeeding at security practitioner exams (e.g., CISM), CTF (Capture The Flag) challenges with and without human-in-the-loop~\cite{shao2024empirical}, etc. The findings suggest these evolving models can narrow the divide between attackers and defenders. 

On the other hand, 
LLMs also introduce significant security and privacy challenges, potentially expanding the attack surface in organizations where LLMs or LLM-integrated applications are deployed. Factors such as the novelty, scale, efficiency, and effectiveness of potential attacks, coupled with the unprecedented growth of new LLM-powered applications, add to the concerns.
However, cyber security stands out as a domain where the concept and practice of red teaming has long been established.  
Now, red teaming is also being performed on LLM models and applications, during the different phases of LLM training, fine-tuning and operation. This evolution encourages a new synergy between ML and security researchers, architects, and engineers.
It is also worth noting that the LLM security domain is witnessing multifaceted activities spanning industry, academia and government bodies, including the development of AI safety frameworks, the formation of alliances, the drafting of regulations, and the definition of processes.
This comprehensive approach holds promise for mitigating LLM security risks and pave way for responsible development in this exciting field.

\bibliographystyle{ieeetr}
\bibliography{ref.bib}

\begin{IEEEbiography}{Dinil Mon Divakaran}
(Senior Member, IEEE; dinil\_divakaran@i2r.a-star.edu.sg) is a Senior Principal Scientist at the A*STAR Institute for Infocomm Research in Singapore. He is also an Adjunct Assistant Professor of the School of Computing, at the National University of Singapore (NUS). 
His research experience and interests include topics such as AI for security, network security and privacy, phishing attacks, as well as endpoint protection. He carried out his doctoral studies at ENS Lyon, France, in the joint lab of INRIA and Bell~Labs.
%He is a Senior Member of the IEEE. 
\end{IEEEbiography}

\begin{IEEEbiography}{Sai Teja Peddinti} (psaiteja@google.com)
is a Staff Research Scientist at Google. His research focuses on applying machine learning/AI and data analysis techniques to build novel privacy and security solutions. He has published papers in many top research venues. He completed his PhD in Computer Science from New York University (NYU), School of Engineering. 

\end{IEEEbiography}

\end{document}